# High-power, mid-infrared, picosecond pulses generated by compression of a CO$_2$ laser beat-wave in GaAs

J.J. Pigeon, S. Ya. Tochitsky and C. Joshi

*Department of Electrical Engineering, University of California at Los Angeles, 405 Hilgard Avenue, Los Angeles, California 90095*
*Corresponding author: jpigeon@ucla.edu*

We report on the generation of a train of ~ 2 ps, 10 µm laser pulses via multiple four-wave mixing and compression of an infrared laser beat-wave propagating in the negative group velocity dispersion region of bulk GaAs and a combination of GaAs and NaCl. The use of a 200 ps, 106 GHz beat-wave, produced by combining laser pulses amplified on the 10P(20) and 10P(16) transition of a CO$_2$ laser, provides a novel method for generating high-power, picosecond, mid-IR laser pulses at a high repetition rate. By using 165 and 882 GHz beat-waves we show that cascaded phase-mismatched difference frequency generation plays a significant role in the four-wave mixing process in GaAs.

OCIS Codes: (140.3470) Lasers, carbon dioxide; (190.4380) Nonlinear optics, four-wave mixing; (320.5520) Pulse compression.

Applications in photonics and strong-field physics have driven a surge of interest in ultrafast, mid-IR laser sources operating at a high repetition rate. Powerful laser pulses with wavelengths ≥ 7 µm will open an opportunity to use highly nonlinear materials such as GaAs and CdTe in the negative group velocity dispersion (GVD) region for the production of broadband IR radiation and ultrafast mid-IR waveforms. GaAs-based generators of coherent white light in the so-called molecular "finger print" region, are already covering this range from 2 to 20 µm [1-4]. Such sources are of interest for spectroscopy and remote sensing of the atmosphere.

Strong-field physics applications for high-power, mid-IR laser pulses arise from the quadratic scaling with laser wavelength of the efficacy of light-matter interaction. For example, mid-IR laser pulses have been successfully applied for shock wave acceleration of monoenergetic protons in a gas jet plasma [5].

Typically optical parametric amplification (OPA) is used to convert the broadband, near-IR light available from solid-state laser systems to produce 0.1 – 1 ps, wavelength-tunable, mid-IR pulses at a high repetition rate [6, 7]. Frequency downconversion of a ~ 1 µm pump in a nonlinear crystal faces significant challenges with an increase of wavelength beyond 5 – 7 µm. As a result, a high-power OPA into the mid-IR relies on a multi-stage frequency conversion and the amplification of a chirped pulse limiting the maximum conversion efficiency and achieved power. Using this approach, 90 GW pulses at a wavelength of 3.9 µm have been produced [8].

The CO$_2$ laser can be used to generate 0.1 - 15 TW, picosecond pulses in the vicinity of 10 µm [9-11]. The relatively narrow bandwidth of this gaseous medium, however, makes picosecond pulse amplification using CO$_2$ challenging. Picosecond CO$_2$ systems [9-11] rely on a broadband, solid-state laser to produce the initial mid-IR seed pulse. This seed is then amplified in a high pressure (<10 atm) [9-11] and/or isotopic laser mix [10] in order to sustain the ~ 1 THz bandwidth required for picosecond pulse amplification. For a high-repetition rate and high energy system, one would like to use a transversely-excited atmospheric (TEA) CO$_2$ laser since, at these pressures, gas-discharges can be operated at 0.1 - 1 kHz, and 1-10 J amplifiers are feasible. The 3.5 GHz bandwidth limitation of gas lasers at 1 atm, however, necessitates a method to convert the long output pulse to a picosecond or shorter pulse duration.

Nonlinear optics provides a broad arsenal of techniques for pulse-length manipulation including self-phase modulation in a nonlinear medium with positive GVD followed by compression, solitonic self-compression in a nonlinear medium with negative GVD and multiple four-wave-mixing compression of a dual frequency beat-signal. The latter technique, whereby a laser beat-wave propagates and evolves in a nonlinear medium, was successfully used to generate trains of transform-limited, 0.1 – 1 ps pulses from an initially cw seed in fibers that exhibit negative GVD at 1.55 µm [12-13]. However, applicability of such a compression technique for the generation of high-power, ultrafast, mid-IR pulses in a bulk nonlinear material has not been explored experimentally. Note that a broadband 8 – 14 µm source based on this principle has been discussed theoretically by Kapetanakos, *et. al.* [14]. In this letter, we experimentally demonstrate the generation of a train of high-power, ~2 ps, 10 µm, laser pulses due to the combination of an increased bandwidth caused by multiple four-wave mixing of a 200 ps long, CO$_2$ laser beat-wave and the effect of negative GVD in GaAs and a combination of GaAs and NaCl.

The laser pulses for this experiment were produced using a master-oscillator power-amplifier (MOPA) system at the UCLA Neptune laboratory which is described in detail elsewhere [11]. The MOPA system is capable of producing 3 - 200 ps long, transform limited, 10 μm laser pulses operating at a pulse repetition frequency of 1 Hz. A pair of 10P(20) and 10P(16) $CO_2$ laser lines, amplified synchronously in the same cavity, were chosen for this study providing a wavelength difference of $\Delta\lambda$ = 10590 nm – 10550 nm = 40 nm or a frequency difference of $\Delta\nu$ = 106 GHz. In this experiment we have used 150 MW, 200 ps long $CO_2$ laser envelope (see Fig. 1) modulated at the 106 GHz beat-frequency with output evenly distributed between the 10P(20) and 10P(16) lines. In experiment the laser fluence was below the 1-1.5 J/cm² damage threshold of the AR coatings of the GaAs samples, which provided nearly 100% transmission.

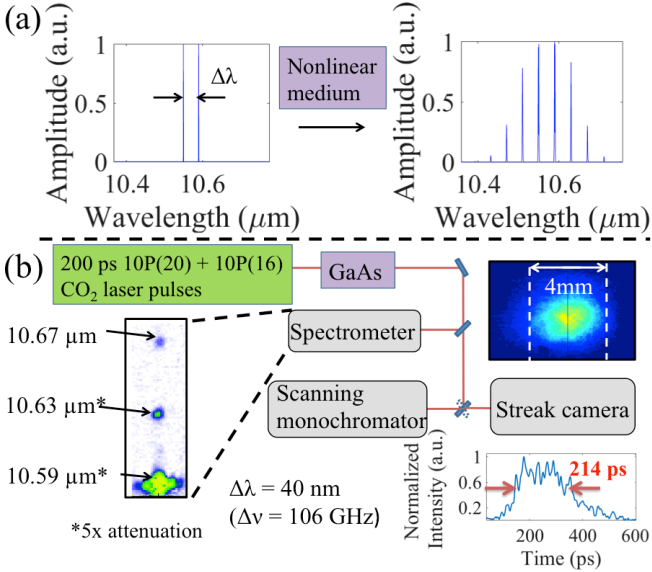

Fig. 1: (a) A dual-wavelength laser pulse produces a comb of sidebands separated by the beat frequency after passing through a nonlinear medium. (b) Schematic of the experimental set-up for the compression of a $CO_2$ laser beat-wave in GaAs. The left insert is a typical spectrum which shows the first two sidebands on the red side after the pulse propagates through the GaAs crystal. The right inserts show the measured spatial beam profile (top) and the temporal envelope of the beat-wave (bottom).

The laser beam, focused to a peak intensity of 0.75 GW/cm², was sent through a single-pass of one or two 67 mm long, semi-insulating, Cr-doped GaAs crystals or in a double pass configuration providing interaction lengths of 67, 134 or 268 mm, respectively.

After the GaAs crystal, the spectrum and temporal pulse structure of the output radiation were measured. As depicted in Fig. 1, the beam is split after the GaAs to perform two separate spectral measurements. The first, a single-shot diagnostic, employed a 5 nm resolution spectrometer in combination with an IR pyroelectric camera to monitor the first two side-bands on the red side of the pump. A typical spectrum of the beat-wave signal, modified after propagating through the GaAs crystal, is shown in the insert of Fig. 1. Energy measurements using the pyroelectric camera allowed for a calibration of the side-band yield on an absolute scale. These measurements were applied for normalization of the data obtained by a second spectral diagnostic, a scanning monochromator in combination with a cryogenically cooled HgCdTe detector. The monochromator, limited in resolution to 20 nm, was used to measure the extent of the spectral broadening.

Time domain measurements were realized by using a picosecond streak camera. For streaking, the $CO_2$ laser pulse was upconverted to the visible range. This frequency upconversion was accomplished by co-propagating the intense mid-IR beam with a visible diode laser beam through a $CS_2$ Kerr cell. When placed between two crossed polarizers, this interaction transcribes the temporal structure from the mid-IR pump to the visible probe [15].

We have simulated the propagation of mid-IR beat-waves in GaAs by solving the 1-D Generalized Nonlinear Schrödinger equation (GNLSE) [16]. Spatial effects, possible free carrier generation at a laser intensity ≥ $10^9$ W/cm², and quadratic nonlinearities, all of which can occur in GaAs, were not included in the model. Due to significant limitations of this model in the case of GaAs, we have used $n_2$ as a fitting parameter. It should be noted that there is a range, $(0.4 – 1.7) \times 10^{-4}$ cm²GW$^{-1}$, of $n_2$ values reported in the literature for GaAs [17, 18]. Simulations were performed using the dispersion relation measured by Skauli, et. al. [19]. We modeled the Raman response of GaAs as an exponentially decaying sinusoid with a frequency of 8.55 THz and a bandwidth of 1 THz, values consistent with Raman fluorescence measurements [20].

In the experiment, we observed a significantly broadened spectrum after the laser beat-wave passes through GaAs. The broadened spectrum consisted of a family of discrete sidebands separated by the beat frequency of 106 GHz. The results are shown in Fig. 2 for GaAs lengths of 67, 134 and 268 mm (Fig. 2a, 2b, and 2c respectively). The vertical lines on Fig. 2 correspond to the maximum amplitude of the side-bands as measured with the monochromator. Here we have measured the amplitude of the side-bands until the signal-to-noise ratio reached 2. In Fig. 2 we have also plotted a simulated spectrum obtained by solving the GNLSE for the same interaction lengths (Fig. 2d, 2e, and 2f). As can be seen in Fig. 2c, the spectrum attains a maximum width of ~ 1.76 THz after propagating through 134 mm of GaAs. In the experiment, we observe a striking saturation of the spectral broadening as increasing the interaction length to 268 mm (see Fig. 2e) did not produce additional broadening.

Figure 3 depicts the temporal structure of the beat-wave after propagating in 67, 134 and 268 mm of GaAs (corresponding to the spectra measured in Fig. 2a, 2b and 2c, respectively). These temporal measurements show an approximately 9 ps modulation of the laser due to the 106 GHz beat-frequency for all lengths. To analyze the change in the individual pulses, we measured the average and standard deviation of the FWHM pulse length, obtained by fitting 5 – 10 pulses from the same train with a sech² pulse shape. It should be noted that these pulses, having duration close to the resolution limit of this diagnostic, must be corrected for the rotational response time of $CS_2$ and the instrumental function of the streak camera. The presented pulse lengths were therefore deconvoluted by $\tau^2 = \tau_{meas}^2 - \tau_r^2 - \tau_{ins}^2$ where $\tau_{meas}$ is the measured pulse length, $\tau_r \sim 1$ ps is the rotational response time of $CS_2$ and

$\tau_{ins}$ ~ 1.5 ps is the instrumental function of the streak camera. These measurements, shown in Fig. 3, indicate that the pulse compresses from $\tau$ = 5.7 to 2.3 ps in 268 mm of GaAs. The temporal profile after 67 mm of GaAs (see Fig. 3a) was indistinguishable to the temporal profile without GaAs. As with the spectral data, we have shown simulated temporal pulse profiles for the same lengths of GaAs (Fig. 3d, 3e and 3f) for comparison.

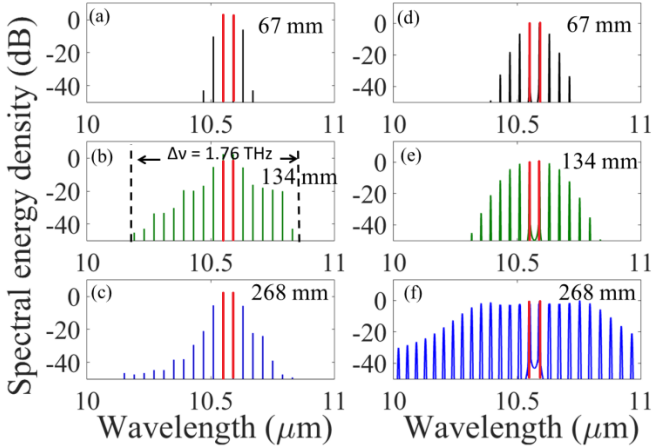

Fig. 2: Experimentally measured (a), (b), (c) and the simulated spectra (d), (e), and (f) after the passage of a 0.75 GW/cm² dual-wavelength, 200 ps laser pulse through different lengths of GaAs. The initial pump wavelengths are shown in red.

The simulation results presented in Fig. 2 and 3 qualitatively match the experimentally measured spectrum and temporal pulse profile for the 67 and 134 mm long interaction lengths (Fig. 2a – 2b and Fig 3a – 3b). Note that the discrepancy between the simulated and experimental side-band yield indicates limitations of the model. The 1-D GNLSE predicts reasonably well the total extent of the side-band production along with pulse compression in the time-domain using $n_2^{eff}$ = $1.7 \times 10^{-4}$ cm²GW⁻¹. Comparison between simulation and experiment for the 268 mm long interaction length, however, indicate a qualitative discrepancy between the simulation and experiment. Specifically, experimental observations indicate that the total extent of the spectral broadening saturates after ~ 134 mm of GaAs, while in simulation the spectrum continues to broaden. Consistent with the simulated spectrum, the simulated temporal pulse lengths qualitatively match the measured pulse lengths for the 67 and 134 mm case but the GNLSE predicts much shorter pulses after 268 mm than what is measured in experiment. One cause of this discrepancy for the 268 mm case is self-focusing in the nonlinear medium, an effect that was not included in the model.

For our experimental conditions the ratio of the laser power to the critical power for self-focusing is approximately P/$P_c$ = 500. This, combined with our beam radius, results in a calculated self-focusing focal length [21] of ~ 150 mm, less than the total interaction length used in the experiment. We have analyzed the spatial beam profile after 67 and 134 mm of GaAs by using magnified images of the laser beam with a spatial resolution of 20 μm. No measurable signs of nonlinear self-focusing or beam distortion were observed. However, measurements of the beam after propagating over 268 mm in GaAs were hindered due to the double-pass configuration. Moreover, tighter focusing of a single-line (10.6 μm) beam - reducing the radius by a factor of two - resulted in an observation of the onset of nonlinear self-focusing for the 134 mm long interaction. Thus for the case of the 268 mm long interaction, a possible laser beam-breakup due to filamentation coupled with a meter-scale distance from the GaAs crystal to our spectral diagnostics could manifest in a poor collection efficiency for the most broad-band components of the laser beat-wave.

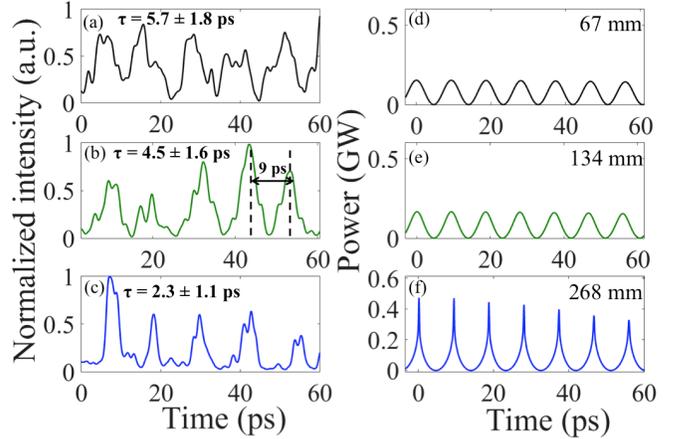

Fig. 3: Temporal pulse structure measured after (a) 67 mm, (b) 134 mm and (c) 268 mm of GaAs. Simulated temporal pulse profiles after (d) 67 mm, (e) 134mm and (f) 268 mm of GaAs.

In order to compress transform-limited pulses it is imperative to balance the effects of nonlinearity and GVD [12]. To achieve a better balance between these two phenomena we replaced the last 134 mm of GaAs with 234 mm of NaCl, a material with a similar GVD as GaAs ($\beta_2$ = -1645 fs²/mm) but with a nonlinear index three orders of magnitude smaller. Figure 4 depicts the measured and simulated temporal pulse profile after the GaAs and NaCl combination. Due to the extra length of NaCl we produced ~1.6 ps pulses, slightly shorter than the 2.3 ps pulses produced with GaAs alone. It should be noted that the measured pulse length is close to the transform-limited pulse length of ~ 1.2 ps.

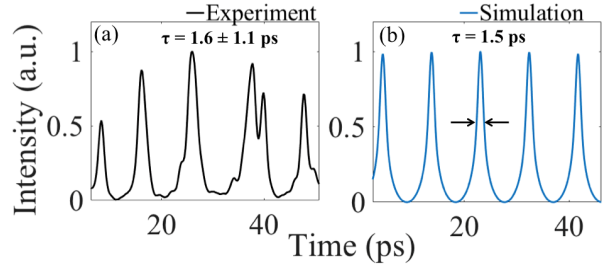

Fig. 4: Experimental (a) and simulated (b) temporal pulse profiles after the beat-wave propagates through 134 mm of GaAs followed by 234 mm of NaCl.

Quadratic nonlinearities, for which GaAs is known, can also be a source of discrepancy between the experimental and simulated spectra. It is well established that phase-mismatched three-wave interactions can lead to an effective third-order nonlinearity $n_2^{eff} = n_2 \pm n_2^{casc}$ where $n_2$ is the Kerr index and $n_2^{casc}$ is the contribution from cascaded quadratic interactions. Here $n_2^{casc}$ is tunable in both magnitude and sign via the phase-mismatch $\Delta k$ [22]. Estimates indicate that $n_2^{casc}$ for second harmonic generation (SHG) provides a small decrease (~10 %) to $n_2^{eff}$ in 10 μm pumped GaAs. Another quadratic interaction that could contribute to $n_2^{casc}$ is difference frequency generation (DFG) between the two pump wavelengths. Room-temperature GaAs has been used to generate THz waves via noncollinear DFG between two IR wavelengths [23].

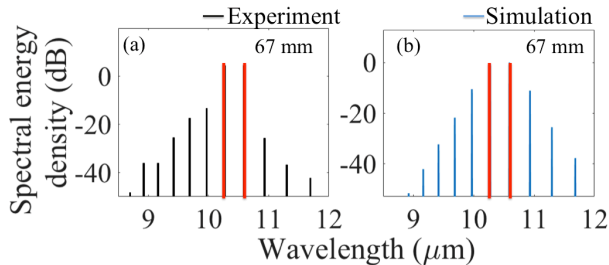

Fig. 5: Experimentally measured (a) and the simulated spectra (b), after the passage of a 0.75 GW/cm$^2$, 882 GHz laser beat-wave through 67 mm of GaAs. The pump wavelengths are colored red.

For the 106 GHz beat-wave, DFG between 10.55 and 10.59 μm is phase-matched ($\Delta k=0$) for collinear propagation due to the anomalous dispersion of the phonon band. To investigate the possible contribution of cascaded DFG we phase-mismatched this interaction by using larger beat frequencies of 165 and 882 GHz. As the beat-frequency was increased we observed increasingly less efficient yield for the first red and blue side-bands. Figure 5a depicts the experimental spectrum after the 882 GHz beat-wave passes through 67 mm of GaAs. For the 882 GHz case ($\Delta k \neq 0$), we needed to simulate the interaction (see Fig. 5b) with $n_2^{eff} = 1.0 \times 10^{-4}$ cm$^2$GW$^{-1}$, a 40% smaller $n_2^{eff}$ than in the case of the 106 GHz ($\Delta k=0$) beat-wave. As can be seen in Fig. 5a, the 882 GHz spectrum has pronounced asymmetry favoring the blue. This interesting observation may be analogous to the asymmetry reported in cascaded SHG [22]. Nevertheless, it is likely that cascading DFG is a universal nonlinear optical phenomenon of mid-IR pumped semiconductors and thus deserves its own detailed study.

To summarize, we have demonstrated a relatively simple method to produce trains of ~2 ps, mid-IR pulses by four-wave mixing of a $CO_2$ laser beat-wave combined with the effect of negative GVD in a GaAs crystal or in a combination of GaAs and NaCl. Since such compression can be done with ~2 ns pulses available from a TEA $CO_2$ MOPA system [24] or a mode-locked TEA $CO_2$ laser, this method can be applied to build compact and affordable short pulse $CO_2$ laser systems giving a train of ps class laser pulses. Finally, by using larger beat-frequencies we have shown that cascaded DFG can greatly affect the nonlinear index of refraction in mid-IR pumped GaAs.

## Acknowledgements

This study was supported by U.S. Department of Energy grant DE-SC0010064.